# Negative-Index Refraction in a Lamellar Composite with Alternating Single Negative Layers

Z. G. Dong, S. N. Zhu, and H. Liu

*National Laboratory of Solid State Microstructures,*

*Nanjing University, Nanjing 210093, China*

Negative-index refraction is achieved in a lamellar composite with epsilon-negative (ENG) and mu-negative (MNG) materials stacked alternatively. Based on the effective medium approximation, simultaneously negative effective permittivity and permeability of such a lamellar composite are obtained theoretically and further proven by full-wave simulations. Consequently, the famous left-handed metamaterial comprising split ring resonators and wires is interpreted as an analogy of such an ENG-MNG lamellar composite. In addition, beyond the effective medium approximation, the propagating field squeezed near the ENG/MNG interface is demonstrated to be left-handed surface waves with backward phase velocity.

# I. INTRODUCTION

Electromagnetic composite metamaterials with simultaneously negative electric permittivity $\varepsilon_{eff}$ and magnetic permeability $\mu_{eff}$, called left-handed materials (LHM), have been a subject of scientific interest over the past several years [1-5]. So far, there are a few ways towards the realization of an LHM. For examples, the metallic metamaterial comprising periodically arranged split ring resonators (SRR) and wires was experimentally confirmed to be an LHM [3]. The possibility of forming an LHM in conducting magnetic granular composite (incorporating conducting ferromagnetic inclusions into an insulating matrix) was theoretically investigated [4]. More recently, Liu et al. proposed a piezoelectric-piezomagnetic multilayer and found that simultaneously negative $\varepsilon_{eff}$ and $\mu_{eff}$ would be available in such a structure [5].

It is easy to understand that, since there is no naturally existing LHM, the straightforward ideas towards the realization of an LHM are how to combine single negative materials, namely, epsilon-negative (ENG) material (where $\varepsilon < 0$, but $\mu > 0$) or mu-negative (MNG) material (where $\mu < 0$, but $\varepsilon > 0$), under appropriate volume fraction to get an effective LHM [3], or alternatively, how to get negative $\mu_{eff}$ in materials having negative $\varepsilon_{eff}$ and vice versa. It was reported that paired structure by ENG and MNG materials can lead to interesting characteristic, such as resonance, tunneling and transparency [6]. Guided mode along the interface of a pairing ENG-MNG waveguide was also theoretically analyzed [7]. An intriguing phenomenon from the alternating layers of ENG and MNG materials is that it can respond effectively in many ways like an LHM, such as, the negative effective index of refraction and the backward phase direction, except that $\varepsilon_{eff}$ and $\mu_{eff}$ do not generally exist [8]. In this paper, we consider such an ENG-MNG lamellar structure to investigate its negative-index refraction by full-wave numerical simulations. It is distinct from Ref. 8 in that the electromagnetic waves in this paper are incident parallel to the laminate slices. In Sec. II, based on effective medium approximation (i.e., the thicknesses of the slices are small enough comparing with the electromagnetic wavelength), negative-index refraction and planar slab focusing with simultaneously negative $\varepsilon_{eff}$ and $\mu_{eff}$ are demonstrated in the structure, and the SRR-wire metamaterials are interpreted as an analogy of the lamellar composite. Section III presents that, beyond the effective medium approximation, the electromagnetic field squeezed near the ENG/MNG interface is in fact left-handed surface waves with negative-index refraction as well as backward phase direction. Finally, conclusions are given in Sec. IV.

## II. EFFECTIVE MEDIUM APPROXIMATION

### A. Left-handed lamellar composite

Consider a lamellar composite as shown in Fig. 1. Two infinite slices, one is an ENG layer ($\varepsilon_1 < 0, \mu_1 > 0$, thickness $d_1$) and the other an MNG layer ($\varepsilon_2 > 0, \mu_2 < 0$, thickness $d_2$), are stacked alternatively along the *x*-axis. Based on the effective medium approximation (i.e., both $d_1$ and $d_2$ are extremely smaller than the electromagnetic wavelength), the components of $\varepsilon$ and $\mu$ can be given by [9-11]

$$\bar{\varepsilon}_x = \frac{\varepsilon_1 \varepsilon_2 (d_1 + d_2)}{\varepsilon_1 d_2 + \varepsilon_2 d_1}, \qquad (1)$$

$$\bar{\varepsilon}_y = \bar{\varepsilon}_z = \frac{\varepsilon_1 d_1 + \varepsilon_2 d_2}{d_1 + d_2}, \qquad (2)$$

$$\bar{\mu}_x = \frac{\mu_1 \mu_2 (d_1 + d_2)}{\mu_1 d_2 + \mu_2 d_1}, \qquad (3)$$

$$\bar{\mu}_y = \bar{\mu}_z = \frac{\mu_1 d_1 + \mu_2 d_2}{d_1 + d_2}. \qquad (4)$$

Therefore, the lamellar structure is a uniaxially anisotropic composite (i.e., it is isotropic in the plane perpendicular to the *x*-axis). For simplicity without loss of generality, suppose $d_1 = d_2$. Then,

$$\bar{\varepsilon}_x \bar{\varepsilon}_y = \bar{\varepsilon}_x \bar{\varepsilon}_z = \varepsilon_1 \varepsilon_2 < 0, \qquad (5)$$

$$\bar{\mu}_x \bar{\mu}_y = \bar{\mu}_x \bar{\mu}_z = \mu_1 \mu_2 < 0. \qquad (6)$$

Such a structure can thus be treated as an indefinite medium in which not all of the principle components of $\varepsilon$ and $\mu$ have the same sign [12]. What is of considerable interest is that, as long as the electromagnetic wave is incident with proper polarization, an indefinite medium (here, the ENG-MNG lamellar composite) may be identical in refraction property with an isotropic LHM [12]. Generally speaking, the prerequisites are as follows:

$$|\varepsilon_1| < |\varepsilon_2|, \text{ and } |\mu_1| < |\mu_2| \quad (\textit{S}\text{-polarized waves}), \qquad (7)$$

$$|\varepsilon_1| > |\varepsilon_2|, \text{ and } |\mu_1| > |\mu_2| \quad (\textit{P}\text{-polarized waves}). \qquad (8)$$

Full-wave numerical simulations are performed to demonstrate the above conclusions. For the sake of simulations, it is not restricted to particular materials. As for the nonmagnetic ENG layer, it can be a metal, metamaterial of intersecting wires, or piezoelectric material, etc., while for the MNG layer; it can be an antiferromagnetic material, SRR metamaterial, or piezomagnetic material, etc.. Consider two slices of an ENG layer ($\varepsilon_1 = -0.4$, $\mu_1 = 0.2$) and an MNG layer ($\varepsilon_2 = 2$, $\mu_2 = -2.2$) with thicknesses $d_1 = d_2 = 0.1$ mm at the given frequency of 50 GHz. According to Eqs. (1)-(4), one gets

$$\bar{\varepsilon} = (\bar{\varepsilon}_x, \bar{\varepsilon}_y, \bar{\varepsilon}_z) = (-1, \ 0.8, \ 0.8) \ , \qquad (9)$$

$$\bar{\mu} = (\bar{\mu}_x, \bar{\mu}_y, \bar{\mu}_z) = (0.44, \ -1, \ -1) \ . \qquad (10)$$

On the one hand, when the electromagnetic wave is *s*-polarized (i.e., electric field in the *x*-direction), as far as the refraction property is concerned, it should be equivalent to an isotropic and homogenous LHM having $\varepsilon = \mu = -1$. As shown in Fig. 2(a), such a lamellar composite in wedge shape is used to simulate its refraction field distribution with appropriate boundary conditions (the surroundings of the wedge is vacuum). From the intuitive viewpoint, the wedge is cut in 45° angle so that the transmitted beam should be bent in 90° angle with respect to the incident *s*-polarized beam [see Fig. 2(a)]. Backward phase moving is also confirmed when the *s*-polarized wave propagates in the wedge, which gives a further evidence of the negative-index refraction in such an ENG-MNG lamellar composite. On the other hand, if the electromagnetic wave is *p*-polarized (i.e., magnetic field in the *x*-direction), it should act as an isotropic medium with $n_{eff} = 0.59$ ($\varepsilon_{eff} = 0.8$, $\mu_{eff} = 0.44$); this is consistent with the numerical result shown in Fig. 2(b). Additionally, from the simulations, it is found the impedance matching condition between the wedge and vacuum is $\eta_{eff} = 1$, which is satisfied in the *s*-polarized case of Fig. 2(a). However, for the *p*-polarized wave, there are obvious reflections due to the impedance of the lamellar composite mismatching to the vacuum. For another example, replacing the slices with an ENG layer $(\varepsilon_1 = -0.5, \mu_1 = 0.5)$ and an MNG layer $(\varepsilon_2 = 2, \mu_2 = -2)$, from which $n_{eff} = -1$ and $\eta_{eff} = 0.75$ for *s*-polarized wave, the reflections at the wedge interfaces are obvious though $|\varepsilon_1|=|\mu_1|$ and $|\varepsilon_2|=|\mu_2|$ in this case (not shown).

Planar slab focusing is another interesting feature of a homogeneous LHM. Though the lamellar composite is effectively an indefinite medium [see Eqs. (9) and (10)], it can still focus *s*-polarized waves as if it were an isotropic LHM due to its in-plane isotropic property ($\mu_y = \mu_z = -1$). In Fig. 3, a current line source excites a cylindrical *s*-polarized wave with a frequency of 50 GHz. As is expected, there are two equidistant images reproduced, one is in the center of the lamellar composite, and the other outside of it. Additionally, strong surface waves are also observed along the interfaces between the lamellar composite and vacuum, which is due to the existence of surface polaritons [13].

### B. Dependence on thickness ratio

For brevity we have confined the discussion under the circumstance of $d_1 = d_2$. In fact, according to Eqs. (1)-(4), it is practical to achieve a certain negative $n_{eff}$, by tailoring the ratio of the thicknesses, without resorting to new materials having different $\varepsilon$ and $\mu$.

### C. Interpretation of the SRR-wire metamaterial

After the demonstration that an ENG-MNG lamellar composite can effectively act as an LHM under the effective medium approximation, we have some remarks on the famous negative refraction experiment by an SRR-wire metamaterial in wedge shape [3]. The original idea was to obtain a metamaterial with its negative $\varepsilon_{eff}$ of the wire array and negative $\mu_{eff}$ of the SRR array, and it was supposed that electromagnetic interactions between the SRR and wire arrays were negligible. However, the interpretation was questioned in Ref. 14, in which the main viewpoint is that an LHM can not be achieved by simply placing wires (negative $\varepsilon_{eff}$) in a homogeneous host with negative $\mu$. From the viewpoint of this work, the SRR-wire metamaterial can be regarded as an analogy of the ENG-MNG lamellar composite, with wire layers as the ENG material and SRR layers as the MNG material [15]. Two relations are given to the SRR-wire metamaterial, no matter it is one-dimensional, two-dimensional, or three-dimensional:

$$\varepsilon_{eff} = \frac{\varepsilon_{SRR} d_{SRR} + \varepsilon_{wire} d_{wire}}{d_{SRR} + d_{wire}}, \qquad (11)$$

$$\mu_{eff} = \frac{\mu_{SRR} \mu_{wire} (d_{SRR} + d_{wire})}{\mu_{SRR} d_{wire} + \mu_{wire} d_{SRR}}, \qquad (12)$$

where $\varepsilon_{SRR}$ and $\varepsilon_{wire}$ are permittivity components in the direction parallel to the wire array, while $\mu_{SRR}$ and $\mu_{wire}$ are permeability components in the direction perpendicular to the SRR-plane. $d_{SRR}$ and $d_{wire}$ are the intervals of SRR and wire layers, respectively.

### III. BEYOND EFFECTIVE MEDIUM APPROXIMATION

In this section, we consider the ENG and MNG materials with thicknesses larger enough than the electromagnetic wavelengths such that the effective medium approximation is not applicable. Because neither the ENG nor MNG material itself supports the propagation of electromagnetic waves, therefore, the transmitted field, if any, is in fact squeezed near the interface layers. The wavevector components in the lamellar composite can be written as

$$k_z^2 - k_{1x}^2 = \varepsilon_1 \mu_1 \frac{\omega^2}{c^2}, \qquad (13)$$

$$k_z^2 - k_{2x}^2 = \varepsilon_2 \mu_2 \frac{\omega^2}{c^2}, \qquad (14)$$

where the transversal wavevector components, $k_{1x}$ and $k_{2x}$, are related as follows [13]:

$$\frac{k_{1x}}{\varepsilon_1} + \frac{k_{2x}}{\varepsilon_2} = 0 \quad \text{(S-polarized waves)}, \qquad (15)$$

$$\frac{k_{1x}}{\mu_1} + \frac{k_{2x}}{\mu_2} = 0 \quad \text{(P-polarized waves)}. \qquad (16)$$

Remind that the thicknesses of the slices are much larger than the wavelengths in this case. Consequently, we obtain

$$k_z^2 = \frac{\varepsilon_1 \varepsilon_2 (\varepsilon_1 \mu_2 - \varepsilon_2 \mu_1)}{\varepsilon_1^2 - \varepsilon_2^2} \frac{\omega^2}{c^2} \quad \text{(S-polarized waves)}, \qquad (17)$$

$$k_z^2 = \frac{\mu_1 \mu_2 (\varepsilon_2 \mu_1 - \varepsilon_1 \mu_2)}{\mu_1^2 - \mu_2^2} \frac{\omega^2}{c^2} \quad \text{(P-polarized waves)}. \qquad (18)$$

Accordingly, effective index of refraction is determined by

$$n_{eff}^2 = \frac{\varepsilon_1 \varepsilon_2 (\varepsilon_1 \mu_2 - \varepsilon_2 \mu_1)}{\varepsilon_1^2 - \varepsilon_2^2} \quad \text{(S-polarized waves)}, \qquad (19)$$

$$n_{eff}^2 = \frac{\mu_1 \mu_2 (\varepsilon_2 \mu_1 - \varepsilon_1 \mu_2)}{\mu_1^2 - \mu_2^2} \quad \text{(P-polarized waves)}. \qquad (20)$$

From Eqs. (19) and (20), one can see that the surface waves transmitted along the ENG/MNG interface has an effective index of refraction. It is interesting and worth of emphasis that the effective index of refraction is negative and hence left-handed surface wave is obtained though no $\varepsilon_{eff}$ and $\mu_{eff}$ can be derived under this configuration; this is confirmed in the following simulations. Consider again the ENG layer ($\varepsilon_1 = -0.4$, $\mu_1 = 0.2$) and MNG layer ($\varepsilon_2 = 2$, $\mu_2 = -2.2$) at frequency of 50 GHz, but now with thicknesses $d_1 = d_2 = 3$ mm (larger thicknesses are less computationally efficient). Firstly, for *s*-polarized waves, negative refraction and anti-parallel phase velocity in the wedge are demonstrated in Fig. 4(a). The negative index from the simulation by using Snell's law is consistent with that in Eq. (19), from where one gets $n_{eff} = -0.315$. Figure 4(b) shows the cross-sectional field distribution. It is found that the electromagnetic field is squeezed near the ENG/MNG interface and degenerated much quickly in the MNG layer than that in the ENG layer. This is because the transversal fields away from the interface are evanescent exponentially into the ENG and MNG materials with the factors of *exp*($-k_{1x}x$) and *exp*[$-k_{2x}(x-d_1)$], respectively [13], where $k_{1x}/k_{2x} = 0.4/2$ from Eq. 15. Secondly, such an ENG/MNG interface can not support the propagation of *p*-polarized waves, since one gets $n_{eff}^2 < 0$ from Eq. 20.

## IV. SUMMARY AND CONCLUSION

In conclusion, negative-index refraction is demonstrated in an ENG-MNG lamellar composite with full-wave simulations on the scenes of the wedge-based negative refraction and planar slab imaging. Firstly, $\varepsilon_{eff}$ and $\mu_{eff}$ of such a lamellar structure can be obtained based on the effective medium approximation, which hence can be regarded effectively as a left-handed composite. Secondly, SRR-wire metamaterials can be interpreted as an analogy to this model. Thirdly, When the ENG and MNG materials are thicker enough than the electromagnetic wavelength; the propagation of the electromagnetic field is demonstrated numerically to be left-handed surface waves with negative effective index of refraction as well as backward phase direction.

## ACKNOWLEDGMENTS

This work was supported by the State Key Program for Basic Research of China (Grant No. 2004CB619003), and by the National Natural Science Foundation of China under Contract No. 90201008.

FIG. 1. Schematic diagram of a lamellar composite with ENG and MNG materials stacked alternatively. Only two periods of the multilayer structure are shown.

FIG. 2. (Color online) The magnetic field magnitudes of wedge-based refractions at 50 GHz incidence. The wedge-shaped lamellar composite is cut in 45° angle with thicknesses $d_1 = d_2 = 0.1$ mm. (a) *S*-polarized wave; (b) *P*-polarized wave.

FIG. 3. (Color online) The magnetic field magnitude of planar slab imaging excited by a current line source at frequency of 50 GHz.

FIG. 4. (Color online) The magnetic field magnitudes of wedge-based refraction at 50 GHz *s*-polarized incidence. The wedge-shaped lamellar composite is cut in 45° angle with thicknesses $d_1 = d_2 = 3$ mm. (a) Longitudinal distribution; (b) Transversal distribution (only one interface was simulated for clarity).

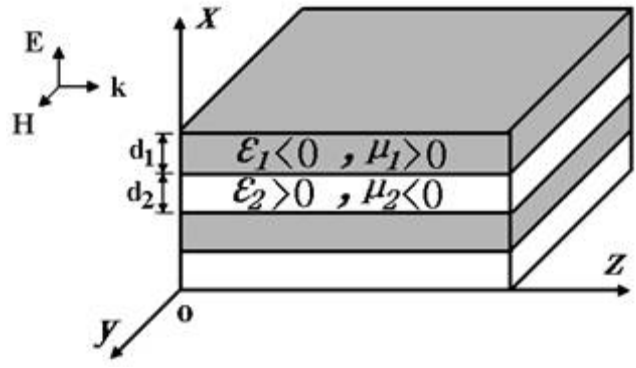

Fig. 1  Z. G. Dong

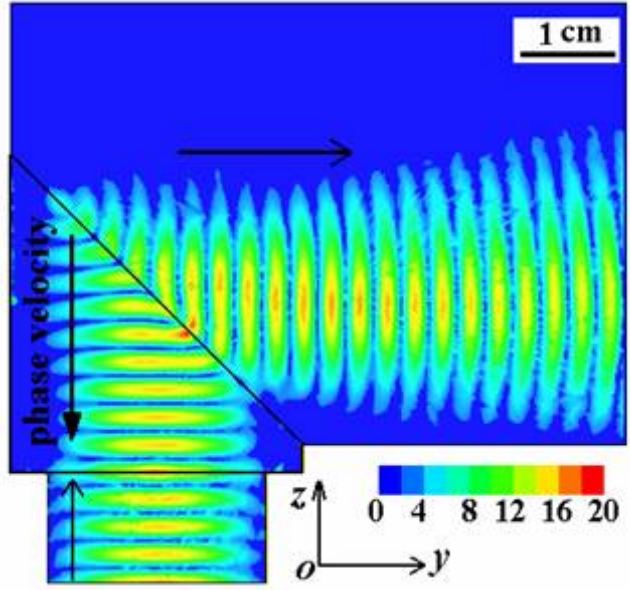

Fig. 2(a)    Z. G. Dong

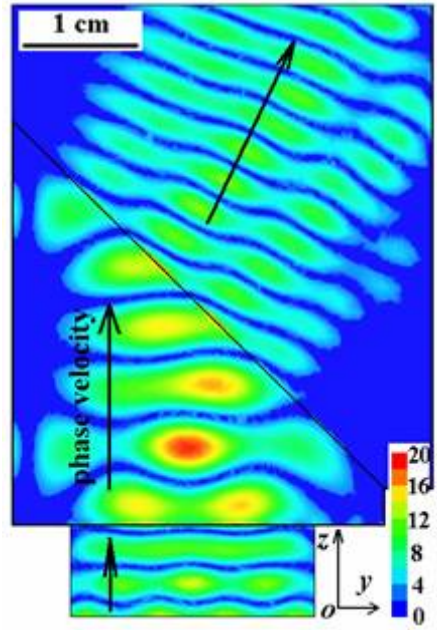

Fig. 2(b)   Z. G. Dong

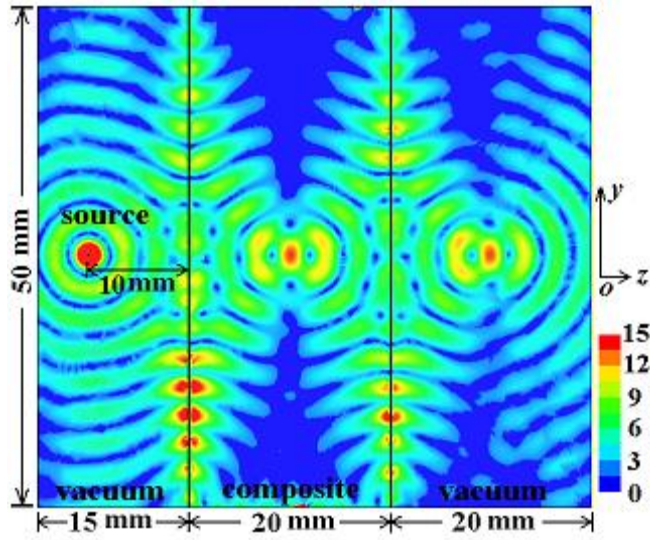

Fig. 3　Z. G. Dong

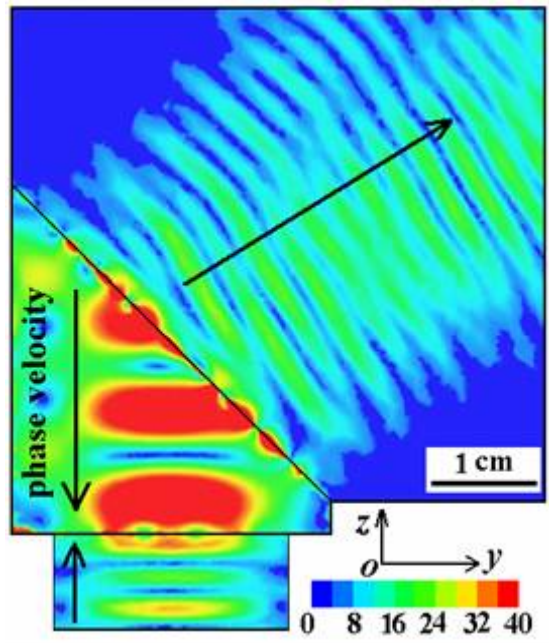

Fig. 4(a)  Z. G. Dong

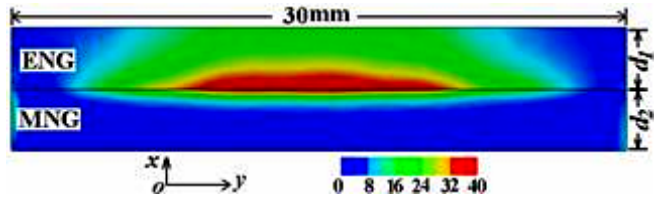

Fig. 4(b)　Z. G. Dong